\documentclass[12pt]{article}
\usepackage{a4wide}
\usepackage{amssymb}
\usepackage{amsmath}
\usepackage{graphicx}
\usepackage{mathdots}
\usepackage{slashed}
\usepackage{verbatim}
\usepackage{xcolor}
\usepackage[utf8]{inputenc}

\usepackage{hyperref}

\usepackage{IEEEtrantools}

\def\makeatletter{\catcode`\@=11}
\makeatletter
\def\mathbox#1{\hbox{$\m@th#1$}}%
\def\math@ccstyles#1#2#3#4#5#6#7{{\leavevmode
      \setbox0\mathbox{#6#7}%
      \setbox2\mathbox{#4#5}%
      \dimen@ #3%
      \baselineskip\z@\lineskiplimit#1\lineskip\z@
      \vbox{\ialign{##\crcr
             \hfil \kern #2\box2 \hfil\crcr
             \noalign{\kern\dimen@}%
             \hfil\box0\hfil\crcr}}}}
\def\mathaccstyles{\math@ccstyles\maxdimen}
\def\maththroughstyles{\math@ccstyles{-\maxdimen}}
\def\unity%
 {\maththroughstyles{.45\ht0}\z@\displaystyle {\mathchar"006C}\displaystyle 1}
 
\usepackage{tikz}
\usetikzlibrary{positioning,arrows}
\usetikzlibrary{decorations.pathmorphing,calc}
\usetikzlibrary{decorations.markings}
\newif\ifmirrorsemicircle
\usepackage{float}
\usepackage{ifthen}
\usepackage{enumerate}
\usepackage{amsmath}
\usepackage{amssymb}
\usepackage[mathscr]{eucal}
\usepackage[errorshow]{tracefnt}
\usepackage{amsmath}
\usepackage{slashed}
\usepackage{amssymb}
\usepackage{array}
\usepackage{amsthm}
\usepackage{eucal}
\usepackage{epsf}
\usepackage{epsfig}
\usepackage{psfrag}
\usepackage{multirow}
\usepackage{wasysym} 
\usepackage{tikz-cd} 
\usepgfmodule{shapes}
\usetikzlibrary{backgrounds, arrows,calc,shapes,decorations.pathreplacing, decorations.markings, automata,positioning}
\usepackage{verbatim}

 \numberwithin{equation}{section}
\begin{document}

\begin{flushright}\footnotesize

\texttt{}
\vspace{0.6cm}
\end{flushright}

\mbox{}
\vspace{0truecm}
\linespread{1.1}

\centerline{\LARGE \bf The large charge limit of scalar field theories}
\medskip

\centerline{\LARGE \bf  and the Wilson-Fisher fixed point at 
$\epsilon=0$ }
\medskip


\vspace{.5cm}

 \centerline{\LARGE \bf }

\vspace{1.5truecm}

\centerline{
    {\bf G. Arias-Tamargo${}^{a,b}$} \footnote{guillermo.arias.tam@gmail.com}
    { \bf D. Rodriguez-Gomez${}^{a,b}$} \footnote{d.rodriguez.gomez@uniovi.es}
    { \bf J. G. Russo ${}^{c,d}$} \footnote{jorge.russo@icrea.cat}}

\vspace{1cm}
\centerline{{\it ${}^a$ Department of Physics, Universidad de Oviedo}} \centerline{{\it C/ Federico Garc\'ia Lorca  18, 33007  Oviedo, Spain}}
\medskip
\centerline{{\it ${}^b$  Instituto Universitario de Ciencias y Tecnolog\'ias Espaciales de Asturias (ICTEA)}}\centerline{{\it C/~de la Independencia 13, 33004 Oviedo, Spain.}}
\medskip
\centerline{{\it ${}^c$ Instituci\'o Catalana de Recerca i Estudis Avan\c{c}ats (ICREA)}} \centerline{{\it Pg.~Lluis Companys, 23, 08010 Barcelona, Spain}}
\medskip
\centerline{{\it ${}^d$ Departament de F\' \i sica Cu\' antica i Astrof\'\i sica and Institut de Ci\`encies del Cosmos}} \centerline{{\it Universitat de Barcelona, Mart\'i Franqu\`es, 1, 08028
Barcelona, Spain }}
\vspace{1cm}

\centerline{\bf ABSTRACT}
\medskip 
We study the sector of large charge operators $\phi^n$ ($\phi$ being the complexified scalar field) in the $O(2)$ Wilson-Fisher fixed point in $4-\epsilon$ dimensions that emerges when the coupling takes the critical value $g\sim \epsilon$. We show that, in the limit 
$g\to 0$, when the theory naively approaches the gaussian fixed point, the sector of operators with $n\to \infty $ at fixed $g\,n^2\equiv \lambda$ remains non-trivial. Surprisingly, one can compute the exact 2-point function and thereby the non-trivial anomalous dimension of the operator $\phi^n$ by a full resummation of  Feynman diagrams. The same result can be reproduced from a saddle point approximation to the path integral, which partly explains the existence of the limit. Finally, we extend these results to the  three-dimensional $O(2)$-symmetric theory with $(\bar{\phi}\,\phi)^3$ potential.

\noindent

\newpage

\tableofcontents

\section{Introduction and summary}

Despite decades of huge research efforts, Quantum Field Theory (QFT) is far from analytic reach beyond perturbative approaches which, in practice, typically comprise the computation of certain observables to a few loop accuracy. It thus comes as a very welcome surprise that, in certain cases, it is possible to identify special limits which lead to drastic simplifications and sometimes to a reorganization of perturbation theory. 
A prototypical example is the large $N$ limit discovered by 't Hooft. In a gauge theory, one takes the rank $N$ of the gauge algebra  to infinity at the same time that the Yang-Mills coupling $g_{\rm YM}$ is sent to zero in such a way that the 't Hooft coupling $g_{\rm YM}^2\,N$ is fixed. This limit selects planar diagrams in the perturbative expansion of the theory, which naturally organize themselves into a genus expansion very reminiscent of a string theory, a connection that has been intensively studied over the last two decades. 

A  different approach is to explore asymptotic regimes in the space of operators in a certain QFT, in particular focusing on those with large charge $n$ under a global symmetry of the theory. This remarkable suggestion was made in \cite{Hellerman:2015nra} 
and it was further explored in
many relevant papers including \cite{Alvarez-Gaume:2016vff,Monin:2016jmo,Hellerman:2017efx,Loukas:2017lof,Banerjee:2017fcx,Hellerman:2017sur,Cuomo:2017vzg,Hellerman:2018xpi,Loukas:2018zjh,Hellerman:2018sjf,delaFuente:2018qwv,Favrod:2018xov,Banerjee:2019jpw,Orlando:2019hte}. A new perturbation expansion emerges in terms
of a small effective coupling represented by the inverse of the charge, $1/n$ (see also \cite{Kravec:2018qnu,Kravec:2019djc} for other interesting physical applications).

Very recently, a new ``double-scaling" large charge limit was introduced in \cite{Bourget:2018obm} in the context of ${\cal N}=2$ 4d superconformal field theories.
In this case, supersymmetric localization provides an efficient method to compute
``extremal" correlators of chiral primary operators $({\rm Tr} \phi^2)^n$ \cite{Gerchkovitz:2016gxx} (being $\phi$ the scalar field in the vector multiplet). 
In the ${\cal N}=2$ SCFT context, the double scaling limit of \cite{Bourget:2018obm} corresponds to taking  $g_{\rm YM}\to 0$, $n\to\infty $ keeping $\lambda = g_{\rm YM}^2\,n$ fixed. This limit systematically isolates, at each loop order in the perturbative expansion of SQCD, a certain contribution. Its existence requires that at $k$ loops, any extremal correlator has a leading behavior $n^k$, which remarkably  turns out to be the case to all loop orders.
Detailed aspects of this limit were discussed in the relevant papers
\cite{Beccaria:2018xxl,Beccaria:2018owt}.
It was recently understood in an important paper  \cite{Grassi:2019txd} that this limit
can be viewed as the standard 't Hooft limit of an associated random matrix
model. In particular, this explains why the limit exists, at least in this theory.
In addition, the matrix model interpretation of  \cite{Grassi:2019txd}  allows one to obtain the exact $\lambda $
dependence in correlators in closed form by employing standard matrix model techniques.

An obvious question is  whether the existence of the double-scaling, large charge limit is a peculiarity of  highly supersymmetric theories such as $\mathcal{N}=2$ SCFTs. In this note we find that an analogous limit exists for a familiar non-supersymmetric theory, namely
scalar field theory with quartic potential.
 We will show that the very familiar Wilson-Fisher (WF) fixed point for the $O(2)$ theory
 provides perhaps the simplest example 
 where one can study non-trivial correlation functions in said limit, by means of a complete resummation of Feynman diagrams.\footnote{The question on the large charge limit of the WF fixed point was also raised in \cite{Grassi:2019txd}, which appeared as this paper was being completed.} 

One may more generally consider the $O(N)$ model in $4-\epsilon$ dimensions, but for
simplicity we shall  restrict the discussion to the $N=2$ case. This can be recast as the theory for a complex scalar $\phi$ with a quartic interaction controlled by a coupling $g$. Appropriately tuning the mass parameter, there is a renormalization group flow to the  Wilson-Fisher fixed point where $g\sim \epsilon$. One of the remarkable applications of the $\epsilon$ expansion is to extrapolate the results to $\epsilon=1$, where the model describes the ferromagnetic transition of the 3d Ising model.  Although this limit is far from the perturbative regime, the analytical results nevertheless remarkably agree with the numerical values for various critical exponents. On the other hand,  in taking the limit $\epsilon\rightarrow 0$   the theory is simply led to the gaussian fixed point in $d=4$. 
However, the limit of \cite{Bourget:2018obm,Beccaria:2018xxl,Beccaria:2018owt,Grassi:2019txd} suggests that
one can consider sectors of large global charge which might have  non-trivial dynamics.
 Specifically, we  consider operators  $\mathcal{O}_n\equiv \phi^n$ of $U(1)$ charge $n$ and engineering dimension $n\,(1-\frac{\epsilon}{2})$. It turns out that in the limit $g\rightarrow 0$, the sector of operators with $n\rightarrow \infty$ 
 such that $\lambda=g\,n^2$ is fixed, have non-trivial correlators, which can be exactly computed through a resummation of the surviving Feynman diagrams. We also provide an alternative derivation from the path integral: in the double scaling limit, it is dominated
 by a saddle-point, giving rise to the same correlation function previously obtained diagrammatically. The saddle-point calculation  suggests that a similar limit may exist in other theories.
 In particular, we also consider the $O(2)$ theory in three dimensions  for a potential $(\bar\phi \phi)^3$, where we identify the relevant limit and compute the
exact two-point correlation function for the operators $\phi^n$, $\bar\phi^n$.

Let us comment on some interesting open problems.
It would be interesting to consider  higher point functions in detail.
A preliminary observation is as follows.
Consider, for instance,  a 3-point function $\langle \mathcal{O}_n(x)\,\mathcal{O}_n(y)\,\bar{\mathcal{O}_{2n}}(0)\rangle$ in the simplest context of the $O(2)$ model studied in this note.  One can show that, to next-to-leading order, there are diagrams surviving the limit, yielding a result consistent with the structure dictated by conformal symmetry. 
Clearly, it would be of interest to extend this study to all orders and to arbitrary $k$-point functions. It would also be very interesting to systematically study   the structure of 2-point functions following \cite{Hellerman:2017sur}. This might lead to universal relations involving the central charges of the conformal algebra. A challenging problem is to see if, as suggested in \cite{Grassi:2019txd}, the double scaling limit 
of the $O(2)$ theory can be understood as a 't Hooft limit of a ``dual" random matrix model.
One may also study large R charge correlators in ABJM theory in the same limit, which could be compared against results from supersymmetric localization.

\section{The Wilson-Fisher fixed point for a complex scalar field}

Let us consider the $O(N)$ model in $d=4-\epsilon$ dimensions. This model is both of pedagogical interest --as the historic laboratory for QFT and RG-- as well as of practical interest: for different values of $N$ it is known to describe various phase transitions of relevant physical systems (for instance, for $N=1$, at $\epsilon=1$, it describes the 3d ferromagnetic transition. See \textit{e.g.} \cite{Simons} for an introduction). The action reads 

\begin{equation}
\label{O(N)}
    S=\int d^{4-\epsilon}x\, \left(\frac{1}{2}\,(\partial\vec{\varphi})^2-\frac{1}{2}\,m^2\,\vec{\varphi}^2-\frac{g}{16}\,(\vec{\varphi}^2)^2\right)\,;
\end{equation}
where $\vec{\varphi}$ is the $N$ component field rotated by the $O(N)$ symmetry. As it is well-known (see \textit{e.g.} \cite{Zinn-Justin}), upon tuning the mass to zero this flows to the 
Wilson-Fisher fixed point at the critical value

\begin{equation}
    g_{\rm WF}=\frac{32\,\pi^2}{N+8}\,\epsilon\,.
\end{equation}

We will be  interested in theories with a global $U(1)$ charge, for which the simplest example is $N=2$. In that case the theory can be re-written as the theory for a complex scalar field in $4-\epsilon$ dimensions with action
\begin{equation}
S=\int d^{4-\epsilon}x\,\left(\partial\bar \phi\,\partial\phi- m^2\,\bar \phi\,\phi-\frac{g}{4}\,(\bar\phi \phi)^2\right)\,.
\end{equation}
With these conventions,  the Feynman rule for the vertex is just $-i\,g$. We will be interested in the critical case where $m^2=0$.

Note that this construction allows one to take the $g\rightarrow 0$ limit along a family of Conformal Field Theories. Nevertheless, since we are ultimately interested in the extreme weak coupling limit, we may alternatively simply consider the $g\,(\bar{\phi}\,\phi)^2$ theory in $d=4$.



\medskip

It is easy to compute the anomalous dimension of
scalar operators of the form $\mathcal{O}_n=\phi^n$ to $O(g)$.
One finds $\gamma_{\mathcal{O}_n}\sim g\,n^2\sim \epsilon\,n^2$ (see \textit{e.g.} \cite{Rychkov:2015naa}). The emergence of the combination $\lambda=g\,n^2\sim \epsilon\,n^2$
suggests the existence of a double scaling limit:

\begin{equation}
\label{limit}
    g\rightarrow 0\,,\qquad n\rightarrow \infty\,,\qquad \lambda=g\,n^2\,\,{\rm fixed}\,.
\end{equation}
The existence of the limit may also be suggested
by earlier investigations on the exponentiation property of multiparticle amplitudes
  \cite{Libanov:1994ug,Libanov:1995gh,Son:1995wz}. 

\section{The double scaling limit on correlation functions}

We will now investigate the  limit \eqref{limit}
in the exact two-point correlation function $\langle \mathcal{O}_n(x) \bar {\mathcal{O}}_{n'}(0)\rangle$, with $\mathcal{O}_n=\phi^n$, $\bar {\mathcal{O}}_n=\bar \phi^n$ in $d=4-\epsilon$ dimensions. These operators have a definite $U(1)$ charge $n$ and hence they are automatically orthogonal for different charges.\footnote{It should be noted that, at fixed charge $n$, the most general operators are of the form $\mathcal{O}_{n,k}=(\bar{\phi}\phi)^k\,\mathcal{O}_n$. We will restrict to the lowest tower with $k=0$.}

\subsection{Diagrammatic computation}

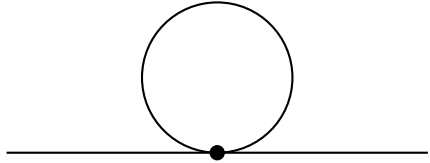
\begin{figure}[h!]
\centering
\begin{tikzpicture}
\draw[thick] (-2.8,0)--(2.8,0);
\draw[thick] (0,1) circle (1);
\draw[thick,fill] (0,0) circle (2.5pt);
\end{tikzpicture}
\caption{Bubble diagram.}
\label{bubble}
\end{figure}

Let us first compute the 2-point functions by evaluation of the relevant Feynman diagrams. As a preliminary step, let us consider the bubble diagram in fig.\ref{bubble}, which is ubiquitous in the perturbative expansion of such correlators.

This diagram has no dependence on the external momenta; therefore it can only be proportional to the mass. Since we will be interested in the critical theory, this diagram vanishes. Thus, when computing $\mathcal{O}_n$ correlators, we shall only consider diagrams that do not contain any bubble.

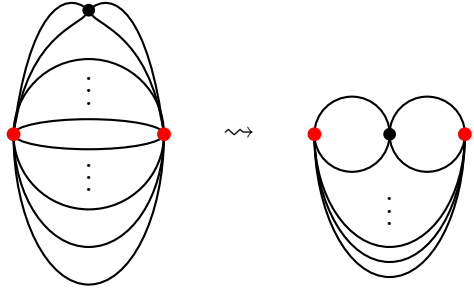
\begin{figure}[h!]
\centering
\begin{tikzpicture}

\draw[thick] (2,0) circle (1);
\draw[thick] (3,0) arc (0:-180:1 and 1.5);
\draw[thick] (3,0) arc (0:-180:1 and 2);
\draw[thick] (2,0) ellipse (1 and 0.2);
\node at (2,0.7) {$\vdots$};
\node at (2,-0.45) {$\vdots$};
\draw[thick] (1,0) .. controls (1.2,1.6) and (1.66,1.95) ..  (2,1.65);
\draw[thick] (1,0) .. controls (1.2,1.4) and (1.95,1.44) ..  (2,1.65);
\draw[thick] (3,0) .. controls (2.8,1.4) and (2.05,1.44) ..  (2,1.65);
\draw[thick] (3,0) .. controls (2.8,1.6) and (2.33,1.95) ..  (2,1.65);
\draw[thick,fill] (2,1.65) circle (2pt);
\draw[thick,fill,red] (3,0) circle (2.2pt);
\draw[thick,fill,red] (1,0) circle (2.2pt);

\node at (4,0) {$\rightsquigarrow$};

\draw[thick] (5.5,0) circle (0.5);
\draw[thick] (6.5,0) circle (0.5);
\draw[thick] (7,0) arc (0:-180:1 and 1.5);
\draw[thick] (7,0) arc (0:-180:1 and 1.7);
\draw[thick] (7,0) arc (0:-180:1 and 1.9);
\draw[thick,fill,red] (5,0) circle (2.2pt);
\draw[thick,fill,red] (7,0) circle (2.2pt);
\draw[thick,fill] (6,0) circle (2pt);
\node at (6,-0.9) {$\vdots$};

\end{tikzpicture}
\caption{The relevant one-loop diagram and its \textit{Kermit the frog} representation. 
}
\label{alien}
\end{figure}

Let us now consider the systematics of the perturbative expansion 
of the correlation function $\langle \mathcal{O}_n(x)\,\bar{\mathcal{O}}_n(0)\rangle$.
As usual, at each order in the perturbative expansion in $g$ there are several topologically different diagrams, each one coming with a certain dependence on $n$.
We are going to be interested in taking $n$ to infinity, and inspection of all topologies shows that in this limit a class of them dominates over the rest.
As shown below, the dominant topology can be viewed as an iteration of the one-loop diagram
of fig. \ref{alien}, that we will call {\it Kermit the frog}'s diagram \cite{kermit}.


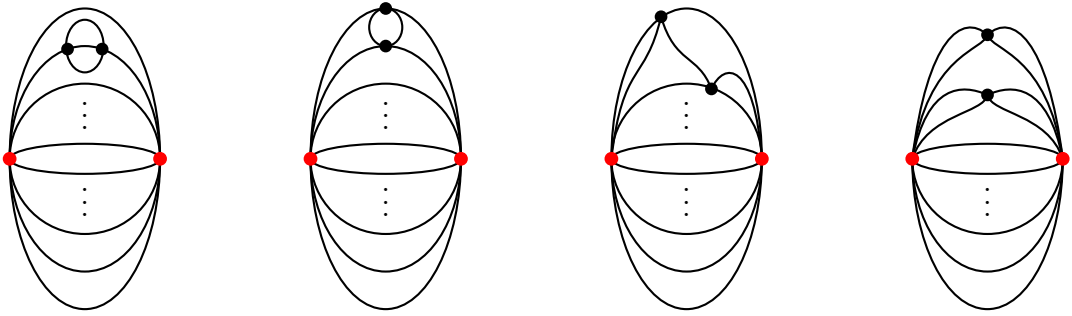
\begin{figure}[h!]
\centering
\begin{tikzpicture}

\draw[thick] (-2,0) circle (1);
\draw[thick] (-2,0) ellipse (1 and 1.5);
\draw[thick] (-2,0) ellipse (1 and 2);
\draw[thick] (-2,0) ellipse (1 and 0.2);
\node at (-2,0.7) {$\vdots$};
\node at (-2,-0.45) {$\vdots$};
\draw[thick] (-2,1.5) ellipse (0.25 and 0.35);
\draw[thick,fill] (-2.23,1.46) circle (2pt);
\draw[thick,fill] (-1.77,1.46) circle (2pt);
\draw[thick,fill,red] (-3,0) circle (2.2pt);
\draw[thick,fill,red] (-1,0) circle (2.2pt);

\draw[thick] (2,0) circle (1);
\draw[thick] (2,0) ellipse (1 and 1.5);
\draw[thick] (2,0) ellipse (1 and 2);
\draw[thick] (2,0) ellipse (1 and 0.2);
\node at (2,0.7) {$\vdots$};
\node at (2,-0.45) {$\vdots$};
\draw[thick] (2,1.75) ellipse (0.22 and 0.25);
\draw[thick,fill] (2,1.5) circle (2pt);
\draw[thick,fill] (2,2) circle (2pt);
\draw[thick,fill,red] (3,0) circle (2.2pt);
\draw[thick,fill,red] (1,0) circle (2.2pt);

\draw[thick] (6,0) ellipse (1 and 2);
\draw[thick] (6,0) ellipse (1 and 0.2);
\draw[thick] (6,0) circle (1);
\node at (6,0.7) {$\vdots$};
\node at (6,-0.45) {$\vdots$};
\draw[thick] (7,0) arc (0:-180:1 and 1.5);
\draw[thick] (5,0) .. controls (5.15,1.3) and (5.45,1) .. (5.66,1.89);
\draw[thick,fill] (5.66,1.89) circle (2pt);
\draw[thick] (7,0) .. controls (6.9,1.3) and (6.5,1.3) .. (6.33, 0.93);
\draw[thick,fill] (6.33,0.93) circle (2pt);
\draw[thick] (5.66,1.89) ..controls (5.85,1.25) and (6.15,1.4) .. (6.33,0.93);
\draw[thick,fill,red] (7,0) circle (2.2pt);
\draw[thick,fill,red] (5,0) circle (2.2pt);

\draw[thick] (10,0) ellipse (1 and 0.2);
\draw[thick] (11,0) arc (0:-180:1);
\node at (10,-0.45) {$\vdots$};
\draw[thick] (11,0) arc (0:-180:1 and 1.5);
\draw[thick] (11,0) arc (0:-180:1 and 2);
\draw[thick] (9,0) .. controls (9.2,1.6) and (9.66,1.95) ..  (10,1.65);
\draw[thick] (9,0) .. controls (9.2,1.4) and (9.95,1.44) ..  (10,1.65);
\draw[thick] (11,0) .. controls (10.8,1.4) and (10.05,1.44) ..  (10,1.65);
\draw[thick] (11,0) .. controls (10.8,1.6) and (10.33,1.95) ..  (10,1.65);
\draw[thick,fill] (10,1.65) circle (2pt);
\draw[thick] (9,0) .. controls (9.2,1) and (9.66,1) ..  (10,0.85);
\draw[thick] (9,0) .. controls (9.2,0.6) and (9.95,0.6) ..  (10,0.85);
\draw[thick] (11,0) .. controls (10.8,0.6) and (10.05,0.6) ..  (10,0.85);
\draw[thick] (11,0) .. controls (10.8,1) and (10.33,1) ..  (10,0.85);
\draw[thick,fill] (10,0.85) circle (2pt);
\draw[thick,fill,red] (9,0) circle (2.2pt);
\draw[thick,fill,red] (11,0) circle (2.2pt);

\end{tikzpicture}
\caption{Four topologies contributing at order $O(g^2)$.}
\label{comparison}
\end{figure}

At order $g^2$, we have the four diagrams of fig. \ref{comparison} above. The key point to identify the dominant diagram at large $n$, at any given loop order, is the $n$ dependence, which comes from the
combinatorial factor. This is given by $n! n!/k!$, where $k$ is the number of lines that do not undergo interactions.
Therefore, the diagram that has the highest power of $n$ is the one
with the smallest $k$.
Using this formula, we thus find that the combinatorial factor of the fourth diagram in fig. \ref{comparison} is $n!\,n\,(n-1)\,(n-2)\,(n-3)$.
On the other hand, the formula $n! n!/k!$ implies that the combinatorial factors of the first, second and third diagrams are $n!\,n $, $n!\,n\,(n-1) $ and $n!\,n (n-1)(n-2)$ respectively (we omit numerical coefficients standing from symmetrization, which do not affect the $n$ dependence). Thus, for large $n$, the  diagram on the right dominates.
We will call this one the {\it two-loop Kermit diagram}.

Now consider the general $m$-loop diagram with $m$ vertices.
The lines in each vertex can go either to another vertex or
join some of the $n$ lines of the operators $\phi^n$ or $\bar \phi^n$. The diagram which has
a smaller number of lines that do not undergo interactions
is when two lines of each vertex join two of the $n$ lines of the operator  $\phi^n$ and the other two lines join two of the $n$ lines of the operator  $\bar \phi^n$ 
(it is not possible to have three lines of the vertex joining 
three lines of the operators $\phi^n$ because of charge conservation; vertices are of the form $\phi\phi\bar\phi\bar\phi$).
This corresponds to the iteration of the Kermit diagram
and has a combinatorial factor $n!^2/(n-2m)!$ which has the
highest power of $n$ (see \eqref{aabb} and below for the  derivation of the combinatorial factor at $m$ loop order including other symmetry factors).

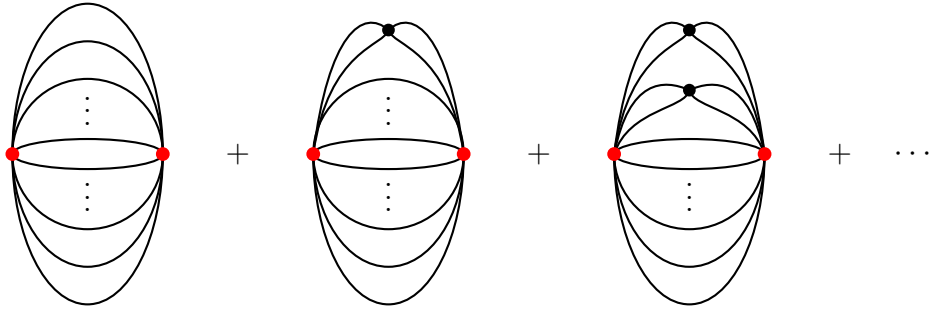
\begin{figure}[h!]
\centering
\begin{tikzpicture}

\draw[thick] (-2,0) circle (1);
\draw[thick] (-2,0) ellipse (1 and 1.5);
\draw[thick] (-2,0) ellipse (1 and 2);
\draw[thick] (-2,0) ellipse (1 and 0.2);
\node at (-2,0.7) {$\vdots$};
\node at (-2,-0.45) {$\vdots$};
\draw[thick,fill,red] (-3,0) circle (2.2pt);
\draw[thick,fill,red] (-1,0) circle (2.2pt);

\node at (0,0) {$+$};

\draw[thick] (2,0) circle (1);
\draw[thick] (3,0) arc (0:-180:1 and 1.5);
\draw[thick] (3,0) arc (0:-180:1 and 2);
\draw[thick] (2,0) ellipse (1 and 0.2);
\node at (2,0.7) {$\vdots$};
\node at (2,-0.45) {$\vdots$};
\draw[thick] (1,0) .. controls (1.2,1.6) and (1.66,1.95) ..  (2,1.65);
\draw[thick] (1,0) .. controls (1.2,1.4) and (1.95,1.44) ..  (2,1.65);
\draw[thick] (3,0) .. controls (2.8,1.4) and (2.05,1.44) ..  (2,1.65);
\draw[thick] (3,0) .. controls (2.8,1.6) and (2.33,1.95) ..  (2,1.65);
\draw[thick,fill] (2,1.65) circle (2pt);
\draw[thick,fill,red] (3,0) circle (2.2pt);
\draw[thick,fill,red] (1,0) circle (2.2pt);

\node at (4,0) {$+$};

\draw[thick] (6,0) ellipse (1 and 0.2);
\draw[thick] (7,0) arc (0:-180:1);
\node at (6,-0.45) {$\vdots$};
\draw[thick] (7,0) arc (0:-180:1 and 1.5);
\draw[thick] (7,0) arc (0:-180:1 and 2);
\draw[thick] (5,0) .. controls (5.2,1.6) and (5.66,1.95) ..  (6,1.65);
\draw[thick] (5,0) .. controls (5.2,1.4) and (5.95,1.44) ..  (6,1.65);
\draw[thick] (7,0) .. controls (6.8,1.4) and (6.05,1.44) ..  (6,1.65);
\draw[thick] (7,0) .. controls (6.8,1.6) and (6.33,1.95) ..  (6,1.65);
\draw[thick,fill] (6,1.65) circle (2pt);
\draw[thick] (5,0) .. controls (5.2,1) and (5.66,1) ..  (6,0.85);
\draw[thick] (5,0) .. controls (5.2,0.6) and (5.95,0.6) ..  (6,0.85);
\draw[thick] (7,0) .. controls (6.8,0.6) and (6.05,0.6) ..  (6,0.85);
\draw[thick] (7,0) .. controls (6.8,1) and (6.33,1) ..  (6,0.85);
\draw[thick,fill] (6,0.85) circle (2pt);
\draw[thick,fill,red] (5,0) circle (2.2pt);
\draw[thick,fill,red] (7,0) circle (2.2pt);

\node at (8,0) {$+$};
\node at (9,0) {$\cdots$};

\end{tikzpicture}
\caption{Diagrams contributing to $\langle \mathcal{O}_n(x)\,\bar{\mathcal{O}}_n(x)\rangle$ at large $n$.}
\label{diagrams}
\end{figure}

Thus, we conclude that a class of diagrams dominate the correlation function, the $m$-loop Kermit diagrams of  fig. \ref{diagrams}. Denoting by $K_m$ the contribution from the Kermit diagram with $m$ interaction vertices, the correlator is

\begin{equation}
\label{aabb}
\langle \mathcal{O}_n(x)\,\bar{\mathcal{O}}_n(0)\rangle = n!\,\sum_{m=0}(-ig)^m\,K_m\,\frac{1}{4^m}\,\frac{n!}{(n-2m)!}\,\frac{1}{m!}\,.
\end{equation}
The combinatorial factor can be understood from fig. \ref{diagrams}.
There are $n$ lines on each side, giving rise to a factor $(n!)^2$ obtained by permutations.
Then one must divide over the number of permutations that lead to equivalent
configurations. There is a factor $1/(n-2m)!$ associated with the permutations
of the $n-2m$ lines that do not undergo interaction.
There is also a factor $1/2^m$ on each side associated with the permutations
of the pair of lines in the $m$ loops. The factor $1/m!$ originates from
the expansion of the exponential of the interaction term.

Using the de Moivre-Stirling formula, for $n\gg1$ one obtains
\begin{equation}
\frac{n!}{(n-2m)!}\approx n^{2m}\ ,\qquad n\gg 1\ .
\end{equation}
Therefore we can define the limit
\begin{equation}
n\rightarrow \infty\,,\qquad g\rightarrow 0\,,\qquad \lambda=g\,n^2={\rm fixed}\,.
\end{equation}
The correlator then becomes

\begin{equation}
\langle \mathcal{O}_n(x)\,\bar{\mathcal{O}}_n(0)\rangle =n!\, \sum_{m=0}K_m\,\Big(\frac{-i\,\lambda}{4}\Big)^m\,\frac{1}{m!}\,.
\end{equation}
To further proceed, note that, in position space, the Kermit diagram $K_m$ is

\begin{eqnarray}
K_m &=& G(0,\,x)^{n-2m}\,\prod_{i=1}^{m}\int d^4z_i\, G(0,\,z_i)^2\,G(z_i,\,x)^2
\nonumber \\ 
&=& G(0,\,x)^{n}\,\Big(\frac{1}{G(0,\,x)^2}\,\int d^4z\, G(0,\,z)^2\,G(z,\,x)^2\Big)^m\,,
\end{eqnarray}
where $G(x,\,y)$ is the propagator of the $\phi$ field. Thus

\begin{equation}
\langle \mathcal{O}_n(x)\,\bar{\mathcal{O}}_n(0)\rangle =n!\,G(0,\,x)^{n}\, \sum_{m=0}\Big(\frac{-i\,\lambda\,\mathcal{K}}{4}\Big)^m\,\frac{1}{m!}\,,
\end{equation}
with
\begin{equation}
\mathcal{K}=\frac{1}{G(0,\,x)^2}\,\int d^4z\, G(0,\,z)^2\,G(z,\,x)^2\,.
\end{equation}
Since $n!\,G(0,\,x)^{n}=\langle \mathcal{O}_n(x)\,\mathcal{O}_n(0)\rangle_0$ is the correlation function in the free theory, and the sum can be trivially resumed, we find

\begin{equation}
\langle \mathcal{O}_n(x)\,\bar{\mathcal{O}}_n(0)\rangle =\langle \mathcal{O}_n(x)\,\bar{\mathcal{O}}_n(0)\rangle_0\, e^{-i\,\frac{\lambda\,\mathcal{K}}{4}}\,.
\end{equation}
Next, consider the computation of $\mathcal{K}$, which is carried out in the appendix. Note that $\mathcal{K}$ represents the $O(g)$ correction to the $\mathcal{O}_2$ correlator. 
We have 

\begin{equation}
\mathcal{K}=-\frac{i}{8\,\pi^2}\,\log(\Lambda^2 x^2)\,.
\end{equation}
As a cross-check of this result, one can see that in the $N=1$ case, and upon appropriately taking into account numerical conventions, this yields the correct $O(g)$ anomalous dimension of the $\mathcal{O}_n$ operator (\textit{cf.}  for example \cite{Rychkov:2015naa}).

Thus

\begin{equation}
\langle \mathcal{O}_n(x)\,\bar{\mathcal{O}}_n(0)\rangle =\langle \mathcal{O}_n(x)\,\bar{\mathcal{O}}_n(0)\rangle_0\,\frac{1}{|x|^\frac{\lambda}{16\,\pi^2}}\,.
\end{equation}
Since in position space

\begin{equation}
G(0,\,x)=\frac{1}{4\,\pi^2}\,\frac{1}{|x|^2}\,,
\end{equation}
we finally find

\begin{equation}
\label{finalOO}
\langle \mathcal{O}_n(x)\,\bar{\mathcal{O}}_n(0)\rangle =\frac{n!}{(4\,\pi^2)^n\,|x|^{2\,(n+\frac{\lambda}{32\,\pi^2})}}\,.
\end{equation}
In particular, this gives the following formula for the dimension of the $\mathcal{O}_n$ operator in the double scaling limit
\begin{equation}
\Delta_{\mathcal{O}_n}=n+\frac{\lambda}{32\,\pi^2}\,.
\end{equation}


\subsection{Saddle-point derivation}

The underlying reason behind the existence of a large charge limit
can be understood from a saddle-point calculation.
It is convenient to rescale the scalar field and
define new variables 
\begin{equation}
    \sigma =g^{\frac14}\, \phi\ ,\qquad    \bar \sigma =g^{\frac14}\, \bar \phi\ .
\end{equation}
The correlator is then given by
\begin{equation}
    \langle \mathcal{O}_n(x_1)\,\bar{\mathcal{O}}_n(x_2)\rangle = \frac{1}{g^{\frac{n}{2}} Z}\int D\sigma D\bar\sigma \ e^{- S}\ ,
\end{equation}
where the Euclidean action, including  source terms,  is given by
\begin{equation}
S=\int d^4 x\,\left(g^{-\frac12} \partial\bar \sigma\,\partial\sigma +\frac{1}{4}\,(\bar\sigma \sigma)^2
-n\delta(x-x_1)\log \sigma -n\delta(x-x_2)\log \bar\sigma  \right)\, .
\end{equation}
In the large $n$ limit, this integral is dominated by a saddle-point.
Indeed, the saddle-point analysis is very similar to the one carried out in section 2.3 of \cite{Hellerman:2017sur}.
The saddle-point equations are given by
\begin{equation}
    \partial^2 \sigma =- ng^{\frac12} \delta(x-x_2)\frac{1}{\bar \sigma}+\frac12 g^{\frac12}\bar\sigma\sigma^2\ ,\qquad
      \partial^2 \bar\sigma =- ng^{\frac12} \delta(x-x_1)\frac{1}{ \sigma}+\frac12 g^{\frac12}\bar\sigma^2\sigma\ .
\end{equation}
The crucial point is that, in the limit $g\to 0$, $n\to\infty$, where $\lambda = n^2 g$ = fixed, the interaction term can be ignored.
The resulting equations become
\begin{equation}
   \bar\sigma  \partial^2 \sigma = - \lambda^{\frac12} \delta(x-x_2)\ ,\qquad \sigma \partial^2 \bar \sigma = - \lambda^{\frac12} \delta(x-x_1)\ .
\end{equation}
These equations  are now equivalent to those discussed in  \cite{Hellerman:2017sur}. The solution is given by
\begin{equation}
\label{solusig}
    \sigma(x) =\lambda^{1/4}\, \frac{e^{i\beta_0}|x_1-x_2|}{2\pi (x-x_2)^2}\ ,
    \qquad   \bar \sigma(x) =\lambda^{1/4}\, \frac{e^{-i\beta_0}|x_1-x_2|}{2\pi (x-x_1)^2}\ .
\end{equation}
Let us now substitute this solution into the action.
Consider first the interaction term. This is absent in \cite{Hellerman:2017sur} and it is indeed  the interesting
part in our case.
We have 

\begin{equation}
    \int d^4 x \frac{1}{4}\,(\bar\sigma \sigma)^2 = \frac{\lambda}{4(2\pi)^4}\int d^4 x \frac{|x_1-x_2|^4}{(x-x_1)^4(x-x_2)^4}\ . 
  \end{equation}
  The integral can be computed by using \eqref{int}, \eqref{Kfinal}, upon shifting $x\rightarrow x+x_1$. We get
\begin{equation}
 \int d^4 x \frac{1}{x^4\,(x-(x_2-x_1))^4}= \frac{4\pi^2}{(x_2-x_1)^4}\,\log|x_2-x_1|\,.
\end{equation}
  Thus
  \begin{equation}
    \int d^4 x \frac{1}{4}\,(\bar\sigma \sigma)^2 = \frac{\lambda}{32\,\pi^2} \log (x_1-x_2)^2\,.
\end{equation}
Let us now consider the remaining terms in the action.
Following \cite{Hellerman:2017sur}, we have
\begin{eqnarray}
  g^{-\frac12}  \int d^4 x\,  \partial\bar \sigma  \,\partial\sigma 
 -  n \log \big(\sigma(x_1) \bar \sigma(x_2)\big) &=& -n\log\big(\sigma(x_1)\bar \sigma(x_2)\big)+n
    \nonumber\\
    &=& -  \frac{n}{2} \log \lambda +n+n\log \left( 2\pi (x_1-x_2)\right)^2 \ .\nonumber
\end{eqnarray}
Putting all pieces together, we  find
\begin{equation}
\label{sfinal}
\langle \mathcal{O}_n(x)\,\bar{\mathcal{O}}_n(0)\rangle =\frac{n!}{(4\,\pi^2)^n\,|x_1-x_2|^{2\,(n+\frac{\lambda}{32\,\pi^2})}}\, \,\qquad n!\sim (2\pi)^{1/2} n^{n+\frac12}e^{-n}\ .
\end{equation}
which is precisely the result \eqref{finalOO} found from the perturbative calculation based on resumming Feynman diagrams.

It is worth noting that \eqref{sfinal} can also be written as

\begin{equation}
\label{finalOObis}
\langle \mathcal{O}_n(x)\,\bar{\mathcal{O}}_n(0)\rangle =\frac{n!}{(4\,\pi^2)^n\,|x|^{2\,n\,(1+\frac{\widehat{\lambda}}{32\,\pi^2})}}\, \ ,
\end{equation}
where $\widehat{\lambda}=\frac{\lambda}{n}=g\,n$. This makes contact with the limit of \cite{Libanov:1994ug,Libanov:1995gh,Son:1995wz}, recently discussed in  \cite{Badel:2019oxl,Watanabe:2019pdh}, where $\widehat{\lambda}$ is kept fixed. More precisely, on general grounds, correlation functions for large charge operators  admit a double, ’t Hooft-like, expansion in $n$, $\widehat{\lambda}$, so that $\Delta = \sum_{i=0}^\infty n^{1-i} F_n(\widehat{\lambda})$. 
At the same time, for weak $\widehat{\lambda}$ coupling, $F_0$ must admit a perturbative expansion $F_0=1+a\,\widehat{\lambda} +\cdots$, where, by explicit computation, $a=\frac{1}{32\pi^2}$. Fixed $\lambda$  implies $\hat\lambda\ll 1$ when $n\gg 1$, which corresponds to the weak coupling regime in the $\widehat{\lambda}$ expansion.
 In this way our formula above is recovered as the $n\rightarrow \infty$ (akin to the planar limit in the familiar ’t Hooft $\frac{1}{N}$ expansion) at weak $\widehat{\lambda}$ coupling. 
From this perspective, the $\frac{1}{n}$ corrections to the saddle-point approximation reconstruct the double expansion described above.

\section{The double-scaling limit in $d=3$}

We can similarly consider $(\bar\phi \phi)^3$ theory in $d=3$, defined by the action

\begin{equation}
S=\int d^{3-\epsilon}x\,\left(\partial\bar \phi\,\partial\phi- m^2\,\bar \phi\,\phi-\frac{g}{3!}\,(\bar\phi \phi)^3\right)\,.
\end{equation}
It should be noted that this model cannot describe the $\epsilon\rightarrow 1$ limit of the WF fixed point discussed in the previous section, as it contains a sextic (as opposed to quartic) interaction. Note that in both cases the interaction term is classically marginal in their respective dimensions and that the fixed points lie in the perturbative regime. In fact, just as in the WF case above, our strategy in this $d=3$ model will be to consider large charge operators in the extreme weak coupling regime.

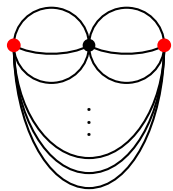
\begin{figure}[h!]
    \centering
    \begin{tikzpicture}
    
\draw[thick] (5.5,0) circle (0.5);
\draw[thick] (6.5,0) circle (0.5);
\draw[thick] (5,0) ..controls (5.25,-0.15) and (5.75,-0.15).. (6,0);
\draw[thick] (6,0) ..controls (6.25,-0.15) and (6.75,-0.15).. (7,0);
\draw[thick] (7,0) arc (0:-180:1 and 1.5);
\draw[thick] (7,0) arc (0:-180:1 and 1.7);
\draw[thick] (7,0) arc (0:-180:1 and 1.9);
\draw[thick,fill,red] (5,0) circle (2.2pt);
\draw[thick,fill,red] (7,0) circle (2.2pt);
\draw[thick,fill] (6,0) circle (2pt);
\node at (6,-0.9) {$\vdots$};

    \end{tikzpicture}
    \caption{Sleeping Kermit. The diagram represents the leading non-trivial contribution
    to the two-point correlation function  of the  $d=3$ theory in the double-scaling limit.}
    \label{fig:sleeppingkermit}
\end{figure}

Let us consider the
saddle-point calculation for the same correlator $  \langle \mathcal{O}_n(x_1)\,\bar{\mathcal{O}}_n(x_2)\rangle$.
After scaling 
$    \sigma =g^{\frac16}\, \phi\ ,\ \     \bar \sigma =g^{\frac16}\, \bar \phi\ .$
the action becomes
\begin{equation}
S=\int d^3 x\,\left(g^{-\frac13} \partial\bar \sigma\,\partial\sigma+\frac{1}{3!^2}\,(\bar\sigma \sigma)^3
-n\delta(x-x_1)\log \sigma -n\delta(x-x_2)\log \bar\sigma  \right)\,.
\end{equation}
A similar saddle-point analysis now leads to
the  equations
\begin{equation}
    \partial^2 \sigma = -ng^{\frac13} \delta(x-x_2)\frac{1}{\bar \sigma}+\frac{1}{12} g^{\frac13}\bar\sigma^2\sigma^3\ ,\qquad
      \partial^2 \bar\sigma = -ng^{\frac13} \delta(x-x_1)\frac{1}{ \sigma}+\frac{1}{12} g^{\frac13}\bar\sigma^3\sigma^2\ .
\end{equation}
We now take the limit
\begin{equation}
    n\to\infty\ ,\ \ g\to 0\ ,\quad {\rm with }\ \ \lambda=n^3g={\rm fixed}\ .
\end{equation}
As in $d=4$, the interaction term vanishes in the limit. The solutions of the saddle-point equations are obtained just like in the $d=4$ case, finding now

\begin{equation}
\label{solusigd=3}
    \sigma(x) =\lambda^{1/6}\, \frac{e^{i\beta_0}|x_1-x_2|^{\frac{1}{2}}}{\sqrt{4\pi} |x-x_2|}\ ,
    \qquad   \bar \sigma(x) =\lambda^{1/6}\, \frac{e^{-i\beta_0}|x_1-x_2|^{\frac{1}{2}}}{\sqrt{4\pi} |x-x_1|}\ .
\end{equation}
The anomalous dimension now comes from the contribution
\begin{equation}
    \int d^3 x \frac{1}{3!^2}\,(\bar\sigma \sigma)^3 = \frac{\lambda}{3!^2(4\pi)^3}\int d^3 x \frac{|x_1-x_2|^3}{|x-x_1|^3\,|x-x_2|^3}\,.
  \end{equation}
This integral represents the Feynman diagram of fig. \ref{fig:sleeppingkermit}, the
``sleeping Kermit".

This integral can be done using the results in appendix, leading to
\begin{align}
      \int d^3 x \frac{1}{3!^2}\,(\bar\sigma \sigma)^3 = \frac{\lambda}{(24\pi)^2}\,\log |x_1-x_2|^2\Lambda^2\,.
\end{align}
The saddle-point calculation implies that, as in the $d=4$ case, this contribution exponentiates, leading to a correlator

\begin{equation}
\langle \mathcal{O}_n(x)\,\bar{\mathcal{O}}_n(0)\rangle =\frac{n!}{( 4\pi)^n\,|x_1-x_2|^{2\,(n+\frac{\lambda}{(24\pi)^2})}}\,,\, \,\qquad n!\sim (2\pi)^{1/2} n^{n+\frac12}e^{-n}\,.
\end{equation}

\subsection*{Note added} After this paper appeared, two papers \cite{Badel:2019oxl,Watanabe:2019pdh} appeared discussing various aspects of the scaling dimension of large charge operators at the WF fixed point in $d=4-\epsilon $. 
The double-scaling limit explored here involves a different regime than the one explored in these works, which in particular permits a full resummation of Feynmann diagrams to all loop orders. As explained below \eqref{sfinal}, we find the expected agreement for the leading term in the weak coupling expansion.

\section*{Acknowledgements}

G.A-T and D.R-G are partially supported by the Spanish government grant MINECO-16-FPA2015-63667-P. They also acknowledge support from the Principado de Asturias through the grant FC-GRUPIN-IDI/2018/000174. G.A-T is supported by the Spanish government scholarship MCIU-19-FPU18/02221. J.G.R. acknowledges financial support from projects 2017-SGR-929, MINECO
grant FPA2016-76005-C.

\begin{appendix}

\section{Real space renormalization}

A relevant integral in our discussion is

\begin{align}
     \mathcal{K}=\frac{1}{G(0,\,x)^2}\,\int d^4z\, G(0,\,z)^2\,G(z,\,x)^2\,,
\end{align}
where the propagator is

\begin{equation}
G(x,\,y)=\frac{1}{4\,\pi^2}\,\frac{1}{(x-y)^2}\,.
\end{equation}

After rotation to euclidean signature, the relevant integral to compute is

\begin{equation}
\label{int}
G(0,\,x)^2\,\mathcal{K}=-\frac{i}{(4\,\pi^2)^4}\int d^4z\,\frac{1}{z^4\,(x-z)^4}\, .
\end{equation}
The integral can be easily computed following the regularization method of \cite{Freedman:1991tk}, i.e. using that, in $d=4$,
\begin{equation}
\label{regulator}
    \frac{1}{z^4}=-\frac{1}{4}\partial^2\left(\frac{\log z^2\,\Lambda^2}{z^2}\right)\,.
\end{equation}
Note that there will be an identical contribution from the divergence at $z=x$, to be regulated just in the same way, and hence the value of the integral will be twice of the contribution at, say $z=0$. We now substitute \eqref{regulator} into the integrand of \eqref{int} and integrate by parts. The resulting integral is convergent upon
giving a small imaginary part to $z$, which does not affect the coefficient of the logarithmic term.
The integral is then  easily computed by going to polar coordinates.
One arrives at
\begin{equation}
\label{Kfinal}
\mathcal{K}=-\frac{i}{8\,\pi^2}\,\log(\Lambda^2\, x^2)\,.
\end{equation}
One may alternatively use the method of  \cite{Chester:2016ref}, which leads to the same result.\\

Next, consider the $d=3$ case. The relevant integral is now

\begin{equation}
\label{int3d}
\int d^3z\,\frac{1}{z^3\,(x-z)^3}\,.
\end{equation}
In this case, it can be regularized using the formula

\begin{align}
    \frac{1}{z^3}=-\frac{1}{2}\partial^2\left(\frac{\log z^2\Lambda^2}{z^2}\right)\,,
\end{align}
and following just the same steps as in the $d=4$ case.

\end{appendix}

\end{document}